\documentclass[10pt,twocolumn,letterpaper]{article}

\usepackage[pagenumbers]{cvpr} 

\usepackage[noend]{algpseudocode}
\usepackage{algorithm}
\usepackage{cuted}
\usepackage{caption}
\usepackage{booktabs}
\usepackage{amssymb}
\usepackage{colortbl}

\usepackage[dvipsnames]{xcolor}

\definecolor{cvprblue}{rgb}{0.21,0.49,0.74}
\usepackage[pagebackref,breaklinks,colorlinks,citecolor=cvprblue]{hyperref}

\def\paperID{*****} 
\def\confName{3DV\xspace}
\def\confYear{2025\xspace}

\title{ReMu: Reconstructing Multi-layer 3D Clothed Human from Image Layers}

\author{Onat Vuran
\and
Hsuan-I Ho
\and
\\
~~~~~~~~~~~~~~~~~~~~Department of Computer Science, ETH Zürich~~~~~~~~~~~~~~~~~~~
\\
{\small\url{https://eth-ait.github.io/ReMu/}}
}

\begin{document}
\twocolumn[{%
\renewcommand\twocolumn[1][]{#1}%
\maketitle
\begin{center}
    \centering
    \captionsetup{type=figure}
    \includegraphics[width=\textwidth]{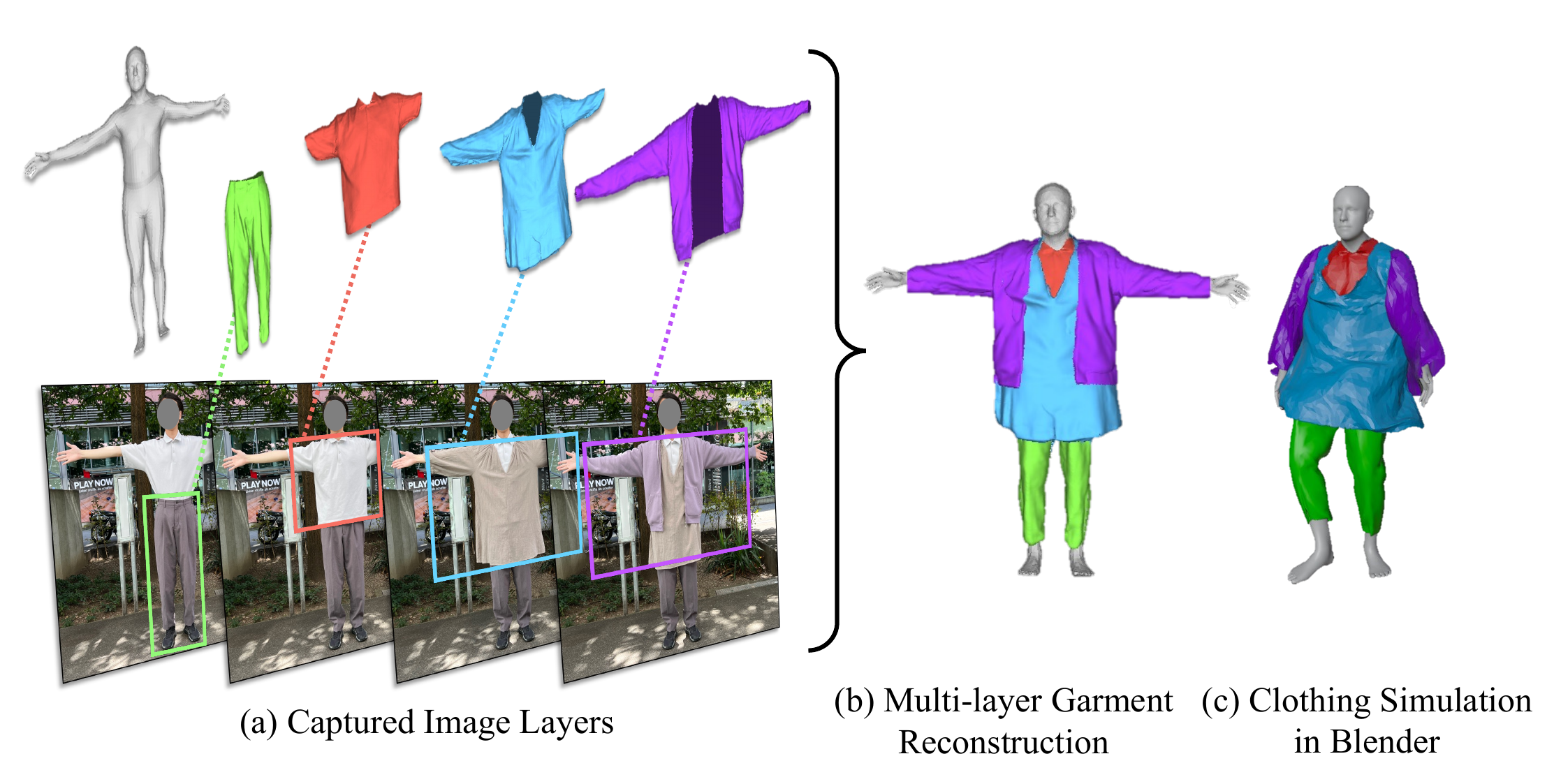}
    \caption{\textbf{ReMu reconstructs multi-layer garments from Image Layers.} (a) Given a set of images captured by a static RGB camera, (b) our method reconstructs 3D garments in layers that fit on a unified human body.
    These layers are optimized to avoid possible inter-layer penetrations, (c) making the resulting 3D garments suitable for various downstream applications, such as clothing simulation.}
    \label{fig:teaser}
\end{center}%
 }]
\begin{abstract}
The reconstruction of multi-layer 3D garments typically requires expensive multi-view capture setups and specialized 3D editing efforts. 
To support the creation of life-like clothed human avatars, we introduce ReMu for reconstructing multi-layer clothed humans in a new setup, Image Layers, which captures a subject wearing different layers of clothing with a single RGB camera.
To reconstruct physically plausible multi-layer 3D garments, a unified 3D representation is necessary to model these garments in a layered manner.
Thus, we first reconstruct and align each garment layer in a shared coordinate system defined by the canonical body pose. 
Afterwards, we introduce a collision-aware optimization process to address interpenetration and further refine the garment boundaries leveraging implicit neural fields. 
It is worth noting that our method is template-free and category-agnostic, which enables the reconstruction of 3D garments in diverse clothing styles.
Through our experiments, we show that our method reconstructs nearly penetration-free 3D clothed humans and achieves competitive performance compared to category-specific methods.
\end{abstract}    
\section{Introduction}
\label{sec:intro}
Digital reconstruction of clothed 3D humans has gained significant attention in academia and industry, across various applications such as virtual try-ons, 3D gaming, and AR/VR telepresence. Numerous approaches have been proposed to reconstruct clothed humans in various setups, including multi-view sequences~\cite{starck2007surface,tsiminaki2014high,zhao2022humannerf,dong2022pina,zheng2024physavatar}, monocular videos~\cite{guo2023vid2avatar,weng_humannerf_2022_cvpr,jiang2022selfrecon,jiang2022neuman,hu2024gaussianavatar,jiang2023instantavatar,chen2021animatable,peng2023implicit,guo2025vid2avatarpro,shin2025canonicalfusion}, and single-view images~\cite{saito2019pifu,saito2020pifuhd,alldieck2022phorhum,xiu2022icon,zheng2021pamir,xiu2023econ,zhang2024sifu,corona2023s3f,han2023high,he2021arch++,liao2023car,huang2020arch,li2024pshuman}. However, these reconstructed 3D models typically represent clothing and the body as a single entity, overlooking the fact that human clothing is often layered. This also limits use cases in traditional graphics pipelines that require separate layers of clothing for applications like physical simulation~\cite{santesteban2022snug,hood,grigorev2024contourcraft} and animation~\cite{lin2024layga}. To overcome this challenge, one needs to reconstruct clothed humans with layers of garments.

Conventionally, reconstructing garments in multiple layers relies on multi-view images to capture respective layers of clothing~\cite{wang20244ddress}. Such a method demands a calibrated multi-view camera system and specialized post-processing for mesh cleanup, which is usually costly and impractical for consumer use. Thus, recent efforts~\cite{jiang2020bcnet,corona2021smplicit,Moon_2022_ECCV_ClothWild,Li2023isp} resort to neural networks to reconstruct multi-layered garment meshes from single-view images. However, it is difficult to see ``beneath" a layer of clothing (e.g., the inside of sleeves under a jacket) from images. To this end, these methods utilize data prior trained with category-specific synthetic data to infer unobserved clothing. Consequently, they often fail to produce realistic results due to the domain gaps between synthetic and real-world clothing. Moreover, we observed that the reconstructed garments may present inter-layer penetrations, making them hard to use for downstream tasks. Therefore, a new technique is required to enable the easy capture of data and the high-fidelity reconstruction of penetration-free garments. 

In this paper, we propose a novel approach to capture and reconstruct 3D clothed humans using a set of single-view \emph{Image Layers} that depict humans in different layers of clothing (see~\cref{fig:teaser} left).
On the one hand, such Image Layers allow for the ease of data capturing using a single, static RGB camera by having a human subject gradually wear each layer of clothing.
This is a significant advantage over multi-view images or 3D meshes, which require a synchronized volumetric capture system. On the other hand, Image Layers offer necessary visual information for 3D reconstruction in unobserved regions. Therefore, the data requirements of category-specific synthetic garments can be alleviated, making our method template-free and generalizable to garment types.

However, reconstructing multi-layer 3D garments from Image Layers is nontrivial, as these images do not guarantee alignment of body pose or the correct order of garments in 3D space.
Simply putting reconstructed garment meshes together cannot yield physically plausible results in 3D space.
To overcome this challenge, we develop a collision-aware clothing reconstruction method operating on a unified layered representation.
We first reconstruct garment meshes using an off-the-shelf single-view human reconstruction model~\cite{ho2024sith} and segment out the clothing using a SAM-based 3D segmentation method~\cite{wang20244ddress}. Next, we align each reconstructed garment mesh with the canonical body using inverse linear blend skinning (LBS). 
To address the inter-layer penetrations of the clothing, we develop a penetration removal process to correct physically implausible garment surfaces. 
Finally, as the resulting meshes may contain distorted topology and fragmented faces caused by noisy 3D segmentation, we further leverage the representative power of neural unsigned distance fields (UDFs) to re-fit and refine the garment meshes with smoother boundaries.
Note that our reconstruction pipeline is training-free and template-free; therefore, it is suitable for the data capturing of diverse 3D clothing.

In extensive experiments, we demonstrate that our new reconstruction scheme using Image Layers is efficient in data capture and superior in reconstruction quality. We show that with the same input conditions, our method outperformed other category-specific and template-based baselines. Moreover, the reconstructed 3D garments are almost penetration-free, making them compatible with graphic tools (e.g., Blender) for simulation and other applications (see~\cref{fig:teaser} right). In summary, our contributions are:

\begin{itemize}[nosep]
    \item a new template-free, category-agnostic clothed human reconstruction scheme using single-view Image Layers. 
    \item a layered clothing representation combining triangle meshes and neural fields to model multi-layer garments in a unified canonical body.
    \item a garment refinement process to achieve collision-free and high-quality reconstructions.
\end{itemize}
\section{Related Work}
\label{sec:related}

\paragraph{Clothed Human Reconstruction.}

The literature on computer vision and graphics has been dedicated to 3D clothed humans for decades. Traditionally, 3D reconstruction utilized multi-view images~\cite{starck2007surface,tsiminaki2014high,zhao2022humannerf,dong2022pina,zheng2024physavatar} captured by calibrated and synchronized volumetric camera systems~\cite{starck2003model,microsoftstage,mpistage,liu2009point,4DHumanOutfit}. In these conventional methods, human bodies are typically represented as 3D textured meshes.
Recently, with the advancement of neural rendering~\cite{tewari2020state,tewari2022advances} and implicit neural fields~\cite{xie2022neural}, a new branch of research that reconstructs clothed humans from monocular videos~\cite{guo2023vid2avatar,weng_humannerf_2022_cvpr,jiang2022selfrecon,jiang2022neuman,hu2024gaussianavatar,jiang2023instantavatar,chen2021animatable,peng2023implicit,guo2025vid2avatarpro,shin2025canonicalfusion} has emerged. In these methods, human bodies are decomposed into a canonical body shape and pose-dependent deformations, which are represented by neural fields or other explicit models~\cite{POP2021,ho2023custom,Zheng2023pointavatar,kerbl3Dgaussians}. These approaches allow for an end-to-end reconstruction leveraging 2D photometric losses and differentiable neural rendering~\cite{mildenhall2020nerf}. 
Another line of research utilizes data-driven models and generative priors to enable reconstruction from a single image~\cite{saito2019pifu,saito2020pifuhd,alldieck2022phorhum,xiu2022icon,zheng2021pamir,xiu2023econ,zhang2024sifu,corona2023s3f,han2023high,he2021arch++,liao2023car,huang2020arch,li2024pshuman}. These methods depend on models trained with 3D data~\cite{tao2021function4d,xavatar,wang20244ddress} to address depth ambiguity and unobserved areas on human bodies. Although the above-mentioned approaches have achieved high-fidelity reconstruction across various data modalities, they typically represent the body and clothing as a single-layer model. 
Our work further extends the reconstruction setting to a multi-layer scheme and jointly tackles multiple input images.

\paragraph{Multi-layer Clothing Reconstruction.}

Compared to single-layer clothed human reconstruction, clothing reconstruction in multi-layer requires a more complex setup to capture individual pieces of garment~\cite{wang20244ddress,zhu2020deep,MGN,SIZER,antic2024close}. This reconstruction process involves parametric body registration~\cite{Smplify,SMPL-X} and human parsing~\cite{yang2023humanparsing} using multi-view images to align and segment 3D garments on a human body. To overcome the difficulty in data capturing, data-driven approaches~\cite{Physics-Inspired,danvevrek2017deepgarment,MGN,jiang2020bcnet,zhu2022registering,qiu2023RECMV,li2024garment,luo2024garverselod} have been proposed to reconstruct 3D garments from single-view images.
Most of these methods are template-based, employing a feed-forward network trained to predict the shape and displacement parameters of each garment template.
Other approaches, such as SMPLicit~\cite{corona2021smplicit}, ClothWild~\cite{Moon_2022_ECCV_ClothWild}, and DrapeNet~\cite{de2023drapenet}, train a generative model for each garment category and fit the latent code to the observed image.
Although these methods can handle multi-layer garments, they rely on category-specific synthetic data, which often lacks the realism needed for accurate 3D reconstruction.
To enhance reconstruction quality, ISP~\cite{Li2023isp} introduces a shape prior represented by sewing patterns, with subsequent work~\cite{li2024garment} integrating deformation priors to address local details. 
However, these data-driven methods do not fully address the ill-posed problem of reconstructing the inner layers of clothing, which are only partially observed in images. In contrast, our reconstruction scheme captures garments from Image Layers, allowing us to apply a template-free and category-agnostic method to reconstruct multi-layer garments. A comparison of different methods is summarized in~\cref{tab:compare}.

\paragraph{Layered Representation for 3D Garments. }
Traditionally, clothing has been modeled using triangle meshes and 2D sewing patterns~\cite{bian2025chatgarment,li2024garment} to fulfill the requirements of the movie, gaming, and fashion industries.
Recently, neural representations~\cite{corona2021smplicit,Moon_2022_ECCV_ClothWild,reloo,kim2024gala,dong2024tela,gong2024laga,wang2024humancoser,lin2024layga} have emerged as a popular alternative for representing clothing in 3D. Unlike traditional methods, these neural representations enable smoother and higher-quality 3D geometry and appearances.
For example, ULNeF~\cite{santesteban2021ulnefs} introduces an approach for managing multi-layer garments with SDFs, ensuring that each layer remains distinct and collision-free. Implicit UDFs~\cite{chibane2020ndf,de2023drapenet} have been proposed to model open-surface and more diverse garment categories. While such implicit representations successfully achieved higher-quality geometry details, handling their physical penetrations became more complicated. 
Thus, our layered representation combines the strengths of both triangle meshes and implicit UDFs, offering an efficient way to address garment penetrations and reconstruct high-quality 3D geometry.

\definecolor{Gray}{gray}{0.85}
\renewcommand{\arraystretch}{1.05}
\begin{table}[t]
    \centering
    \footnotesize
\begin{tabular}{l|c|c|c|c}
\toprule
    Method & \begin{tabular}[c]{@{}c@{}} Single- \\ view \end{tabular} 
           & \begin{tabular}[c]{@{}c@{}} Multi- \\ layer \end{tabular}
           & \begin{tabular}[c]{@{}c@{}} Template- \\ free \end{tabular} 
           & \begin{tabular}[c]{@{}c@{}} Category- \\ agnostic \end{tabular} \\
\midrule
    PIFu~\cite{saito2019pifu} & \checkmark &  & \checkmark &  \checkmark  \\
    4D-DRESS~\cite{wang20244ddress} &  & \checkmark & \checkmark & \checkmark    \\
    BCNet~\cite{jiang2020bcnet} & \checkmark & \checkmark &  &  \\
    SMPLicit~\cite{corona2021smplicit} & \checkmark & \checkmark & \checkmark &  \\
    \bf Ours & \checkmark & \checkmark & \checkmark &  \checkmark \\
\bottomrule
\end{tabular}
\caption{\textbf{Comparison of clothed human reconstruction.} We reconstruct multi-layer 3D garments from single-view Image Layers. Our method does not require pre-defined garment templates or category-specific data for training.}
\label{tab:compare}
\end{table}

\begin{figure*}[t]
  \centering
  \includegraphics[width=\linewidth]{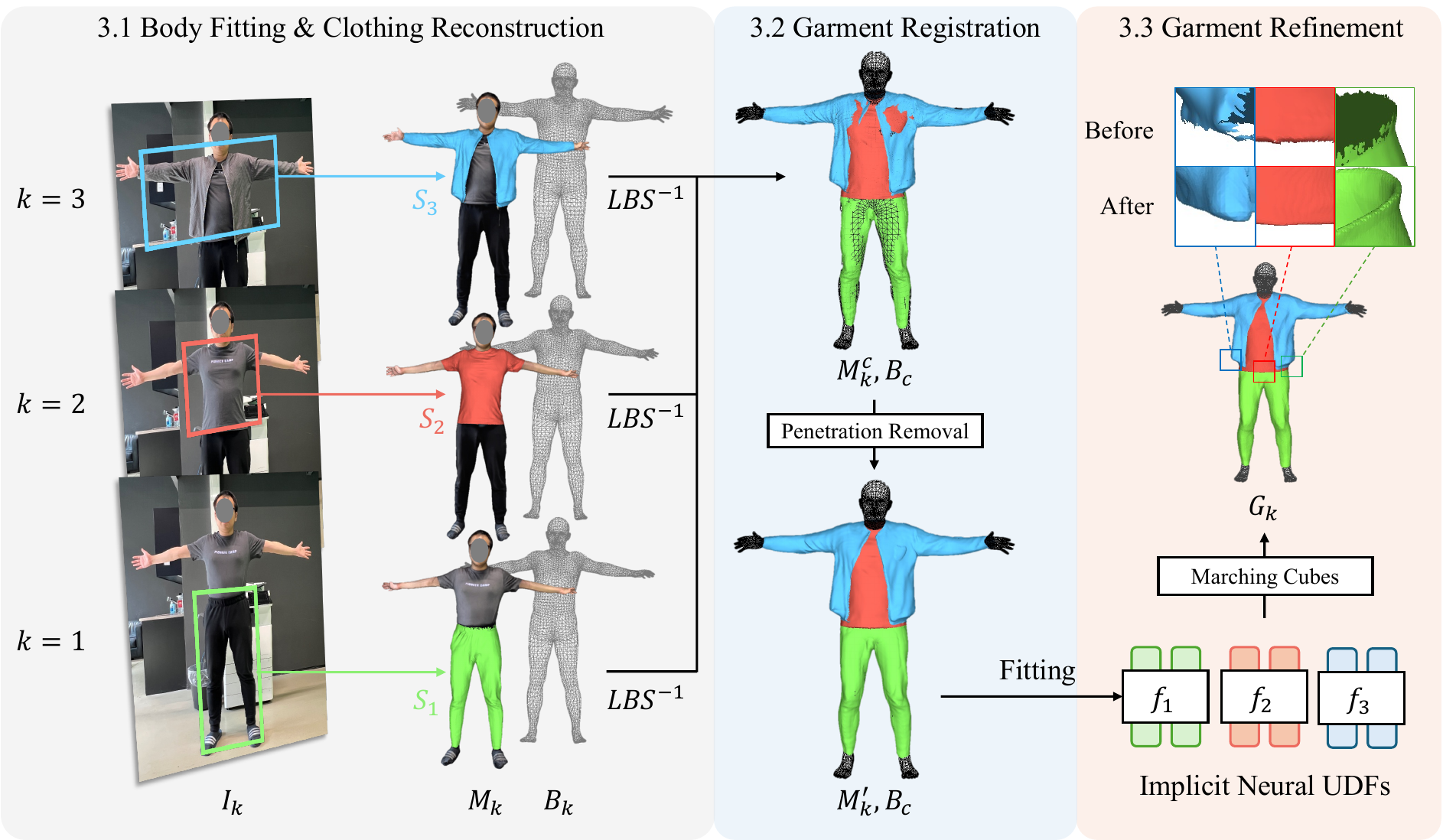}
  \vspace{-1.5em}
  \caption{\textbf{Layered clothing reconstruction pipeline.} 
  Given a set of Image Layers $I$, we register SMPL-X body models $B$ and reconstruct watertight meshes $M$ from images. We then segment out 3D garments $S$ through multi-view parsing (\cref{sec:recon}). Next, we deform 3D meshes to the canonical body pose $B_c$ through inverse LBS. These aligned garments are optimized to remove inter-layer penetrations as $M'$ (\cref{sec:rep}). Finally, we fit implicit neural fields $f$ to refine the garment surface geometry and boundaries (\cref{sec:opt}).
  }
  \label{fig:method}
  \vspace{-1.em}
\end{figure*}

\section{Method}
Given a set of Image Layers $ \{I_1, ... ,I_{N_l}\}$ captured by a single RGB camera, our goal is to reconstruct their corresponding 3D garments $ \{G_1, ... ,G_{N_l}\}$ which are properly dressed on the human subject in layers.
\cref{fig:method} summarizes the workflow of our multi-layer garment reconstruction pipeline. We first fit a parametric body model and reconstruct a single-layer garment mesh at each image layer (\cref{sec:recon}). To register separate pieces of clothing into a shared 3D space, we deform the garment meshes to a canonical space and remove inter-penetrations between layers (\cref{sec:rep}). In achieving high-fidelity 3D reconstruction, we parameterize the garments as implicit UDFs and refine their surface geometry (\cref{sec:opt}).

\subsection{Body Fitting and Clothing Reconstruction}
\label{sec:recon}
For each Image Layer $I_k, k \in \{1, ..., {N_l}\} $, we aim to fit a parametric body model $B_k$ to the image and reconstruct a single-layer watertight mesh $M_k$ of the clothed subject. For body fitting, we employ an SMPL-X estimator~\cite{cai2023smplerx} and refine the estimated body pose by minimizing the 2D keypoint~\cite{openpose} reprojection errors. The resulting body model $B_k$ is then utilized by a single-view 3D human reconstruction model~\cite{ho2024sith} in obtaining a watertight textured mesh $M_k$. However, this textured mesh is composed of multiple garments in a single-layer manner. We need to segment out the clothing of interest at each layer (e.g., the lower garment for $k=1$ in \cref{fig:method} \emph{left}). To this end, we develop a 3D segmentation method similar to the one used in 4D-DRESS~\cite{wang20244ddress} and GALA~\cite{kim2024gala}. We render a set of multi-view images of the textured mesh and use Grounded SAM~\cite{ren2024grounded} to predict 2D segmentation masks of the selected clothing. To obtain a 3D segmentation mask, we reproject each 2D mask to the mesh surface and decide the label of each mesh vertex through voting. Finally, we filter out disconnected vertices and smooth out the vertex labels to prevent holes in the surface, yielding a final clothing area $S_k$ defined by the 3D segmentation labels as shown in \cref{fig:method} \emph{left}.

\subsection{Garment Registration}
\label{sec:rep}

\paragraph{Garment alignment.} The reconstructed 3D garment meshes from the previous step are not aligned in the 3D space due to pose variance in each Image Layer. To properly dress every piece of clothing on top of a unified human body, we define a canonical body pose $B_c$ (e.g., T-pose in \cref{fig:method} \emph{middle}) as the shared 3D coordinate system. We deform each full-body mesh $M_k$ back to the canonical body pose using the pose parameters $B_k$ with inverse linear blend skinning (LBS).
Let $\mathbf{B}_{ki}$ denote the bone transformation matrix for SMPL-X body joint $i \in \{ 1, ..., 21\}$ for pose $B_{k}$, a vertex of the mesh $\mathbf{v}_k \in M_k$ can be represented as:

\begin{equation}
    \mathbf{v}_k = LBS (\mathbf{v}_k^c) = \sum_{i=1}^{21} w_i^c \mathbf{B}_{ki} \mathbf{v}_k^c.
\end{equation}
Therefore, the inverse LBS for $\mathbf{v}_k^c \in M_k^c$ in the canonical pose $B_{c}$ is formulated as:
\begin{equation}
    \mathbf{v}_k^c = LBS^{-1} (\mathbf{v}_k) = (\sum_{i=1}^{21} w_i \mathbf{B}_{ki})^{-1} \mathbf{v}_k
\end{equation}

Note that $M_k^c$ denotes the deformed watertight mesh of layer $k$ in the canonical pose $B_c$, and $w_i, w_i^c$ represent the $i$-th skinning weights
for a vertex in the original pose and the canonical pose, respectively. The skinning weights are queried from the nearest vertex in the SMPL-X body model.

\begin{figure*}[t]
\centering
\includegraphics[width=\linewidth]{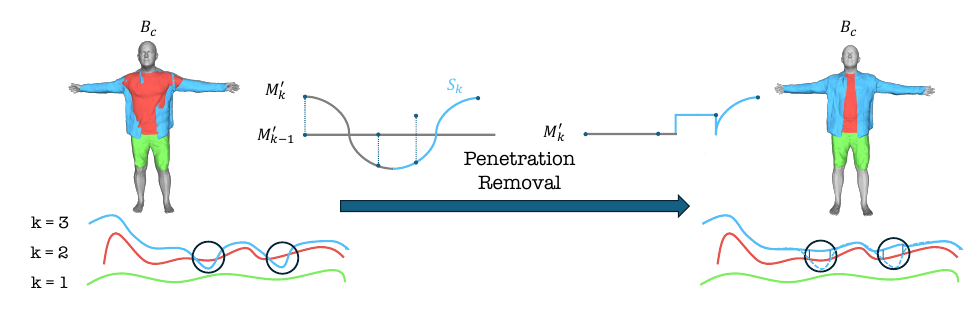}
\caption{\textbf{Visualization of penetration removal}. We iteratively displace the penetrating vertices and register the vertices of non-garment areas. Our algorithm ensures penetration-free garment layers.}
\label{fig:pene}
\end{figure*}

\paragraph{Penetration removal.} While the garment meshes are now aligned on top of the canonical body, there are still potential inter-penetrations across garment layers. To address this, we develop a penetration removal algorithm. By gradually running this algorithm from the inner layer to the outer layer, we align and dress every layer of garments $M_k'$ on the canonical human body $B_c$ and address most garment collisions across layers, as shown in~\cref{fig:method} \emph{middle}.

More specifically, penetration removal aims to modify the vertex locations of $M^c_k$ so that the vertices not belonging to $S_k$ are on the surface of $M^c_{k-1}$, and the vertices belonging to $S_k$ are slightly outside of $M^c_{k-1}$. We represent the output of this procedure with $M'_k$ for $k = 1, \ldots, K$. This ensures that the garment at level $k$ will be more outside the garment at level $k-1$, thus they will be penetration-free. We let $M'_0 = B_c$ and apply the following algorithm.

For each $k \geq 1$, we iterate all the vertices of $M^c_k$. For each vertex $\mathbf{v}^k_j$, we first compute its displacement direction to the body $M'_0$. We define $\mathbf{b}^k_j$ as the nearest point to $\mathbf{v}^k_j$ on the human body surface, and the displacement direction is chosen to be the face normal $\mathbf{n}^k_j$ at the point $\mathbf{b}^k_j$. Then we define the line $\ell^k_j = \mathbf{b}^k_j + t\mathbf{n}^k_j$, parametrized by $t$, and allow the point $\mathbf{v}^k_j$ to be only displaced along this line. The displacement amount is determined differently according to whether $\mathbf{v}^k_j$ belongs to $S_k$ or not.

If the vertex belongs to a non-garment area, i.e., $\mathbf{v}^k_j \notin S_k$, we find the intersection between $\ell^k_j$ and the surface of $M'_{k-1}$ and place the vertex at the intersection point. If the vertex belongs to a garment area, we check whether it is inside the watertight mesh $M'_{k-1}$. If not inside, we leave the vertex as is, not to damage the loose deformations. But if it is inside, it corresponds to an interpenetration case. In this case, we again detect the intersection between $\ell^k_j$ and the surface of $M'_{k-1}$ but place the vertex slightly outside of this intersection point. The amount is decided according to a small thickness parameter $\varepsilon$. 

As a summary,~\cref{fig:pene} illustrates the updating process and~\cref{eq:pene} shows the mathematical expression, where $\mathbf{v}^{k'}_j$ is the displaced vertex and $t^*$ corresponds to the intersection between $\ell^k_j$ and $M'_{k-1}$.

\begin{equation}
\mathbf{v}^{k'}_j = 
\begin{cases}
\mathbf{b}^k_j + (t^* + \varepsilon) \mathbf{n}^k_j, & \text{if } \mathbf{v}^{k'}_j \in S_k \text{ and inside of } M'_{k-1} \\
\mathbf{b}^k_j + t^* \mathbf{n}^k_j, & \text{otherwise}
\end{cases}
\label{eq:pene}
\end{equation}

\subsection{Garment Refinement}
\label{sec:opt}
We observe two major geometric artifacts on these garment meshes after penetration removal (\cref{fig:method} \emph{right}). First, noisy 3D segmentation introduces fragmented mesh topology near the boundaries of the garments. Second, vertex displacements disrupt the original smooth garment surface geometry. To achieve both collision-free and high-quality 3D clothing reconstruction, we employ the representative power of implicit neural fields for garment refinement. We represent $k$-th garment layer using an implicit UDF representation:
\begin{equation}
      f_k : \mathbb{R}^{3} \rightarrow \mathbb{R}_{\geq0}, f_k (\mathbf{x}) = d_k, k\in \{1, ..., {N_l}\}.
\label{eq:udf}
\end{equation}
The value $d_k$ indicates the distance of a point $\mathbf{x}$ to the garment surface $S_k$. We parameterize $f_k$ as a neural network and fit it to the garment by sampling 3D points near the garment surface. The UDF fitting objective is formulated as:
\begin{equation}
    \mathcal{L}_{udf} = \frac{1}{|N_l|} \sum_{k=1}^{N_l} \mathbb{E}_{\mathbf{x}_k\sim S_k}\Bigl[ \left \| f_k(\mathbf{x}_k) - d_k \right \|_{2}^{2}\Bigr] .
\end{equation}
After fitting, the garment meshes $G_k$ can be obtained by Marching Cubes. These garments are smoother and complete in the boundary regions as shown in~\cref{fig:method} \emph{right}.
\definecolor{Gray}{gray}{0.85}
\begin{table*}[t]
\resizebox{\linewidth}{!}{%
    \centering
\begin{tabular}{l|ccc|ccc|ccc}
\toprule
    & \multicolumn{3}{c|}{Lower}          & \multicolumn{3}{c|}{Inner}         & \multicolumn{3}{c}{Outer}          \\
\midrule
 Method               & CD (mm)$\downarrow$ & NC$\uparrow$ & IR (\%)$\downarrow$ & CD (mm)$\downarrow$ & NC$\uparrow$ & IR (\%)$\downarrow$   & CD (mm)$\downarrow$ & NC$\uparrow$ & IR (\%)$\downarrow$    \\
\midrule
 \rowcolor{Gray}SMPLlicit~\cite{corona2021smplicit}        & 13.85 & 0.894 & 0.000 &  23.75 & 0.815 & 0.230 & 31.73 & 0.769 & 6.093  \\
 \rowcolor{Gray}ClothWild~\cite{Moon_2022_ECCV_ClothWild}  & 16.07 & 0.884 & 0.001 & 36.40 & 0.812 & 0.756 & 39.56 & 0.742 & 1.455  \\
 \rowcolor{Gray}ISP~\cite{Li2023isp}                       & 15.56 & 0.894 & 1.522 &  41.68 & 0.825 & 0.185 & 18.62 & 0.813 & 0.140  \\
\midrule
SMPLlicit+                                  & \underline{15.24} & \underline{0.891} & \underline{0.033} &  \underline{20.55} & 0.810 & \underline{0.128} & 31.73 & 0.769 & 5.996  \\
ClothWild+                                  & 16.18 & 0.882 & \textbf{0.001} &  38.87 &  0.794 & 0.136 & 39.56 & 0.742 & 3.575  \\
ISP+                                        & 23.17 & 0.880 & 1.527 &  34.44 & \underline{0.843} & \textbf{0.057} & \underline{18.62} & \textbf{0.813} & \textbf{0.137}  \\ 
Ours                                        & \textbf{8.422} & \textbf{0.927} & 0.052 &  \textbf{14.14} & \textbf{0.855} & 0.204 & \textbf{18.53} & \underline{0.807} & \underline{0.485}  \\
\midrule
GT Reference                    & - & - & 5.762 &  - & - & 6.776 & - & - & 9.997  \\
\bottomrule
\end{tabular}
}%
\vspace{0.3em}
\caption{\textbf{Clothed human reconstruction benchmark}. We computed Chamfer distance (CD), normal consistency (NC), and Intersection Ratio (IR) between ground truth and reconstructed meshes in each category obtained from different baselines.
The best and the second-best methods are highlighted in \textbf{bold} and \underline{underlined} respectively. Note that
the \colorbox{Gray}{gray color} denotes the original methods that reconstruct garments from one image. We expand these methods to utilize Image Layers (denoted as ``+") for fair comparisons. }
\vspace{-1em}
\label{tab:recon}
\end{table*}

\begin{figure*}[t]
\centering
\includegraphics[width=\linewidth]{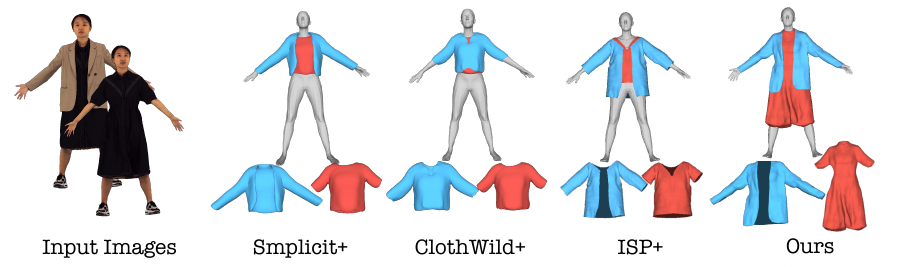}
\vspace{-1.em}
\caption{\textbf{Generalizablity to garment categories}. Existing data-driven methods cannot reconstruct garments of unseen categories in training, such as long coats and dresses. As our method is training-free and template-free, we faithfully reconstructed these garments without any adaptations. }
\vspace{-1.em}
\label{fig:template}
\end{figure*}

\section{Experiments}

\subsection{Experimental Setup}
\paragraph{Dataset.}
We use the 4D-DRESS~\cite{wang20244ddress} dataset to evaluate our method. This dataset contains multi-view images and 3D scans of 32 subjects wearing different layers of clothing, including lower garments (pants, skirts), inner garments (t-shirts, dresses), and outer garments (jackets, coats). We randomly chose 12 subjects to build an evaluation benchmark of multi-layer clothing reconstruction. For each subject, we extract the lower, inner, and outer garment meshes using the ground-truth semantic labels. Since the 3D meshes might be captured in slightly different poses, we deform the lower and inner clothing to the pose of the outer clothing. Finally, we render front-view 2D images as the inputs to the baseline methods. Note that for all the methods, we provide ground-truth poses and labels to rule out errors caused by pose estimation and focus on clothing reconstruction.

\paragraph{Evaluation Metrics.}
We follow the evaluation protocol in single-view mesh reconstruction (e.g., ICON~\cite{xiu2022icon}, SiTH~\cite{ho2024sith}) using Chamfer distance(\textbf{CD}) and normal consistency (\textbf{NC}) to evaluate geometry reconstruction quality. Since inter-penetration is crucial for 3D garments in simulation, we introduce a new metric dubbed Intersection Ratio (\textbf{IR}) to measure inter-penetration. More specifically, we render frontal and back images of garment meshes via rasterization and compute the total observed areas of the garment $A$. Next, we stack all the inner garments together (including the body mesh) and compute again the areas of the outer garment $\hat{A}$ on the rendered images. The Intersection Ratio is calculated as $(A-\hat{A}) / A$, which indicates the percentage of inter-penetration being observed.
\vspace{-1em}
\begin{figure*}[h]
\centering
\includegraphics[width=0.99\linewidth]{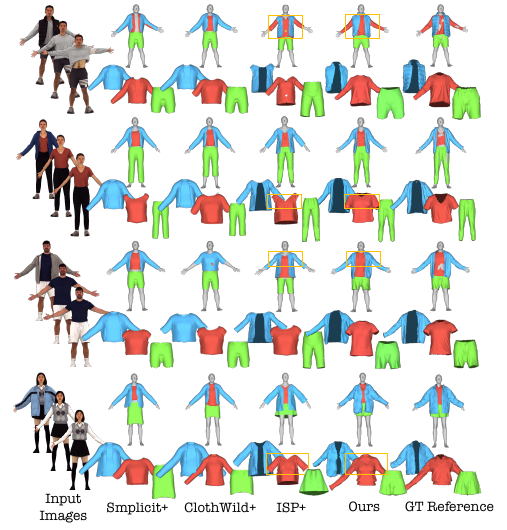}
\caption{\textbf{Multi-layer clothing reconstruction on the 4D-DRESS dataset}. We compare with adapted baseline methods using multiple Image Layers as inputs. The tops are 3D overlays of human bodies and reconstructed garments; the bottoms are reconstructions of each clothing layer.
Existing methods failed to reconstruct the correct garment shapes corresponding to the input images.
Our method produced garments containing local details and matching the ground-truth reference.
Note that the ground-truth garments are aligned with inverse LBS and, therefore, they are not fully penetration-free. 
The garment meshes from SMPLicit, ClothWild, and Ours are double-sided, while ISP and GT Reference are single-sided (not shaded in the open area).
We use SMPL-X as the canonical body for better pose alignment for hands, while other methods use SMPL. Best viewed in color and zoomed in.
}
\label{fig:main}
\end{figure*}

\paragraph{Baselines.}
We compare our method with learning-based clothing reconstruction methods, including
\textbf{SMPLlicit}~\cite{corona2021smplicit}, \textbf{ClothWild}~\cite{Moon_2022_ECCV_ClothWild}, and \textbf{ISP}~\cite{Li2023isp} (\colorbox{Gray}{gray color} in~\cref{tab:recon}). However, these methods only infer garments from a single image (i.e., the outer clothing image) instead of Image Layers. For a fair comparison, we implemented extensions of these methods (denoted with an additional ``+" in ~\cref{tab:recon}) to reconstruct clothing using the same input data as our method. For each Image Layer, we run the methods to obtain the outermost garment mesh and stack all the results on the same canonical body. We also report the IR values of ground-truth scans as a baseline for the multi-view reconstruction method~\cite{wang20244ddress} with inverse LBS.
\\
\\
\begin{table*}[t]
\resizebox{\linewidth}{!}{%

    \centering
    \small
\begin{tabular}{l|ccc|ccc|ccc}
\toprule
    & \multicolumn{3}{c|}{Lower}          & \multicolumn{3}{c|}{Inner}         & \multicolumn{3}{c}{Outer}          \\
\midrule
 Method   &    CD (mm)$\downarrow$ & NC$\uparrow$ & IR (\%)$\downarrow$ & CD (mm)$\downarrow$ & NC$\uparrow$ & IR (\%)$\downarrow$   & CD (mm)$\downarrow$ & NC$\uparrow$ & IR (\%)$\downarrow$    \\
\midrule
 Inverse LBS                                & 9.987 & 0.924 & 17.06 &  17.96 & 0.839 & 35.70 & 24.11 & 0.803 & 27.82  \\
 + Penetration Removal                      & 9.111 & 0.924 & 0.002 &  15.27 & 0.842 & 0.075 & 19.66 & 0.786 & 0.060  \\
 + Garment Refinement                               & 8.422 & 0.927 & 0.052 &  14.14 & 0.855 &  0.204 & 18.53 & 0.807 & 0.485  \\
 \midrule
 GT Reference                      & - & - & 5.762 &  - & - & 6.776 & - & - & 9.997  \\
\bottomrule
\end{tabular}
}
\caption{\textbf{Ablation study on the 4D-DRESS dataset}. We computed Chamfer distance (CD), normal consistency (NC), and Intersection Ratio (IR) between ground truth and the results from each step of our method. Our full model achieved the best 3D metrics but with slightly more penetrations. Note that our overall IR is desirable ($< 0.5 \%$) when compared with multi-view 3D ground truth as a reference.}
\label{tab:ablation}
\end{table*}

\begin{figure*}[t]
\centering
\includegraphics[width=0.9\linewidth]{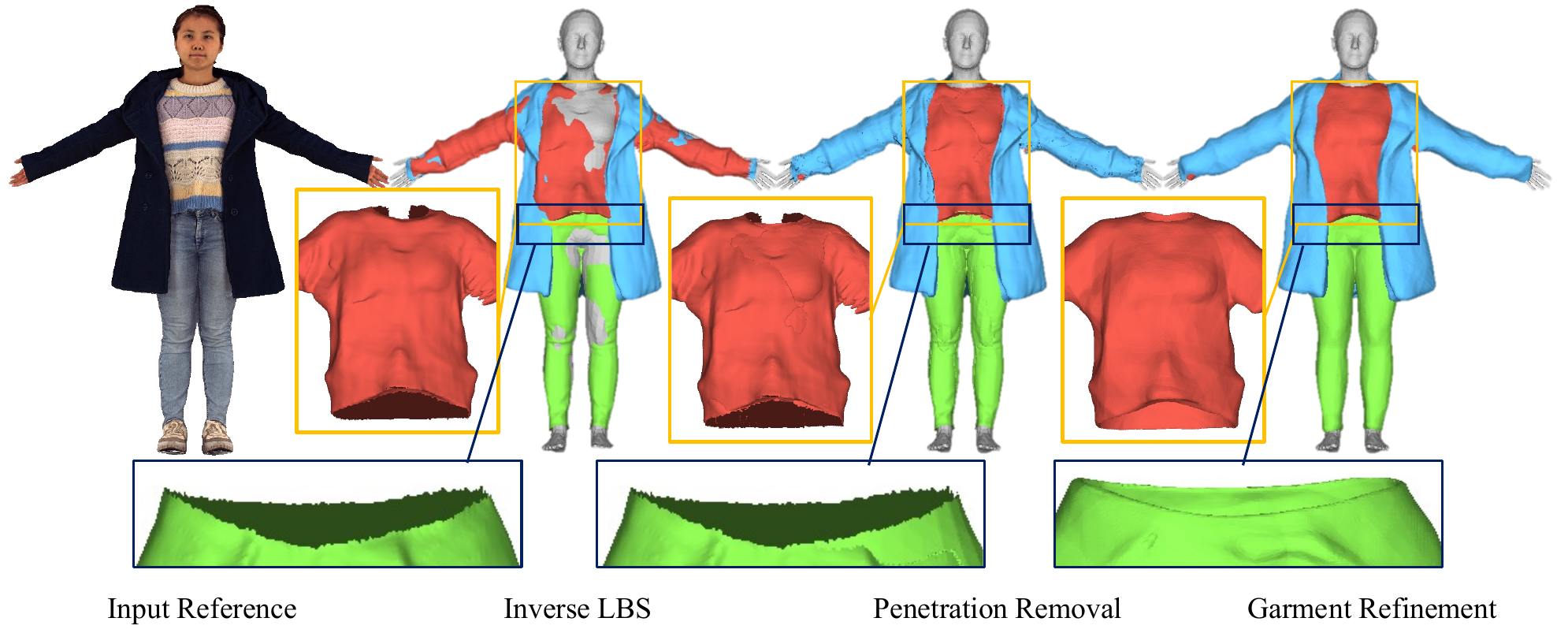}
\caption{\textbf{Visualization of ablation study}. We visualize the results at each step of our method. While penetration removal can address inter-layer penetration problems, the clothing boundaries and geometry are also noisy. With the garment refinement process, we obtain smoother and continuous surfaces by fitting implicit neural fields.}
\vspace{-1.5em}
\label{fig:ablation}
\end{figure*}

\subsection{Multi-layer Garment Reconstruction}
We visualize the qualitative results of multi-layer garment reconstruction in~\cref{fig:main}. Data-driven methods (i.e., SMPLicit, ClothWild) trained with synthetic datasets typically produced inaccurate garment shapes and lacked local details like clothing wrinkles. Although ISP generated more correct garment shapes, it struggled to model proper sleeve lengths and inter-penetrations, as highlighted in the yellow rectangles of~\cref{fig:main}. By contrast, our method reconstructed garment shapes that closely matched the ground-truth scans and tackled the inter-penetration problems in multi-view captured garments. It is worthwhile to mention that our method is generalizable to various garment types, whereas others failed to reconstruct unseen garments in training, such as long coats and dresses in~\cref{fig:template}.

The quantitative comparison results are shown in~\cref{tab:recon}. We see that our method achieved the best results on the 3D reconstruction metrics for both lower and inner clothing. As for the outer garments, our method also presented performance comparable to the SOTA method ISP. Regarding inter-penetrations, our method obtained a desirable ($< 0.5\%$) IR performance in all clothing categories while other methods fell short in either lower or outer garments. 

\subsection{Ablation Study}
We conducted controlled experiments to verify the effectiveness of each component in our methods. The results are summarized in~\cref{tab:ablation} and~\cref{fig:ablation}. As discussed previously, the Penetration Removal process reduced inter-layer penetrations to $< 0.1\%$, but also introduced artifacts on the clothing surfaces. This is evident from the NC metric shown in~\cref{tab:ablation}. To address this, the Garment Refinement step further improved the 3D reconstruction metrics and produced smoother boundaries and surfaces as depicted in~\cref{fig:ablation}. As a tradeoff, this also slightly increased the penetration ratio. Note that this IR value ($< 0.5\%$) remains acceptable for the clothing simulation application using Blender as shown in~\cref{fig:teaser}.

\section{Discussion and Conclusion}
\textbf{Limitation.} 
Our method is currently tailored for human subjects captured in the T/A pose. One exciting future direction is relaxing this requirement by using single-view videos.
Currently, our UDF fitting processing does not incorporate other constraints. Another promising future work is investigating better fitting objectives to ensure both high-quality and collision-free results. Finally, our optimization pipeline has not been optimized for large-scale data collection, which is important for future generative 3D clothing and virtual try-ons.
\paragraph{Conclusion.} We introduce a method for reconstructing multi-layer 3D garments in a new setup, Image Layers.
Our approach, leveraging a layered representation and a penetration-aware optimization process, allows for the template-free, category-agnostic reconstruction of 3D clothing. 
Experimental results demonstrate that our method achieves high-quality results comparable to traditional multi-view setups, simplifying the garment reconstruction process for downstream tasks. 
Our work has the potential to enhance the accessibility of 3D garments, thereby accelerating the future development of generative AI and Metaverse applications.
\paragraph{Acknowledgements.} This work was partially supported by the Swiss SERI Consolidation Grant "AI-PERCEIVE".
{
    \small
    \bibliographystyle{ieeenat_fullname}
    \bibliography{main}
}
\maketitlesupplementary

\begin{strip}
\centering
\includegraphics[width=\linewidth]{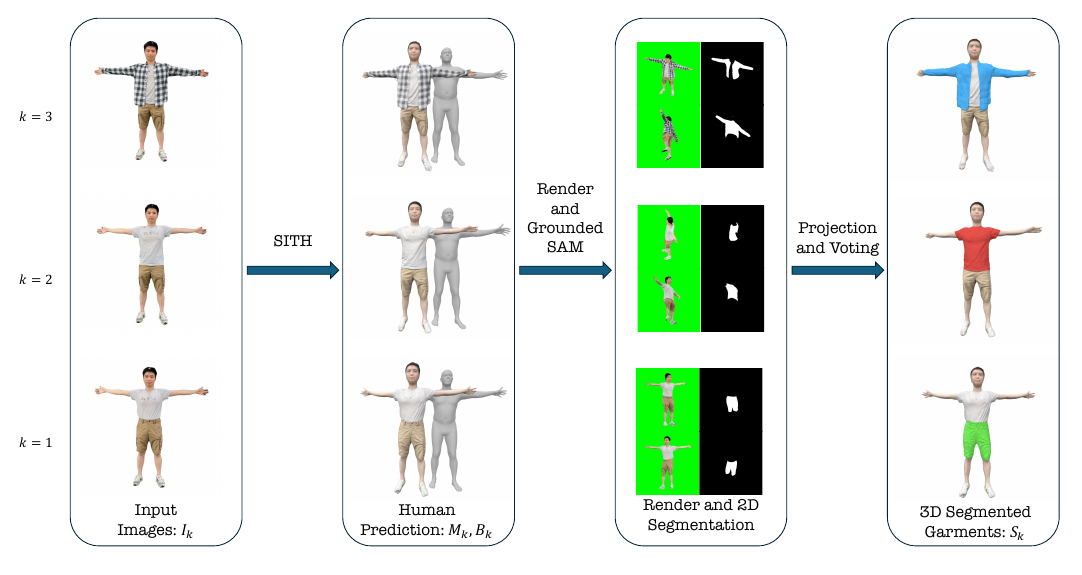}
\captionof{figure}{\textbf{Visualization of body fitting and garment separation}. We utilize SITH to extract a clothed avatar and a human body from the image layers. The respective 3D garment area of the clothed avatars are segmented using 2D segmentations of their multi-view renders by perspective projection and voting. We utilize GroundedSAM for 2D segmentations.}
\label{fig:sep}
\end{strip}

\section{Implementation Details}

\subsection{Body Fitting and Garment Separation}

In the garment reconstruction stage, we first fit a single-layer watertight mesh to each image using SITH~\cite{ho2024sith}. SITH starts by fitting an SMPLX~\cite{SMPL-X} body to the image and then hallucinates the back view of the person in the image using a diffusion model. Finally, it combines image features from front and back views together with the features of the SMPLX fit to reconstruct a watertight 3D clothed avatar. We name the fitted bodies $\{B_1, B_2, ..., B_K\}$ and the reconstructed watertight meshes of clothed avatars $\{M_1, M_2, ..., M_K\}$.

As $\{M_1, M_2, ..., M_K\}$ represents human and garments as a single entity, we use a 3D segmentation method to extract the respective garment area from each mesh. We use a similar approach to GALA~\cite{kim2024gala}. For each mesh $M_i$ in $\{M_1, M_2, ..., M_K\}$, we render a set of multi-view images from which we segment out the respective clothing in 2D using GroundedSAM~\cite{ren2024grounded}. Then, we project all the vertices of $M_i$ to all the views to decide if that vertex belongs to the desired garment via voting. This gives us 3D segments $\{S_1, S_2, ..., S_K\}$ which correspond to the vertices that belong to the desired garment in $\{M_1, M_2, ..., M_K\}$.

While those 3D segments represent very high-quality garments, they can not be separated to be used as garments. They suffer from two major problems. Firstly, they are slightly at different poses and require registration before they can be draped onto a common human body $B_c$. Secondly, they are colliding with each other, namely, the t-shirt is visible on the undesired areas, such as at the back, when draped under the jacket. \cref{fig:sep} depicts examples of those problems. So we apply the later stages to ensure garments are penetration-free and aligned correctly in 3D coordinate space.

\subsection{Fitting Multi-Layered Garment UDFs}

\begin{figure*}[t]
\centering
\includegraphics[width=\linewidth]{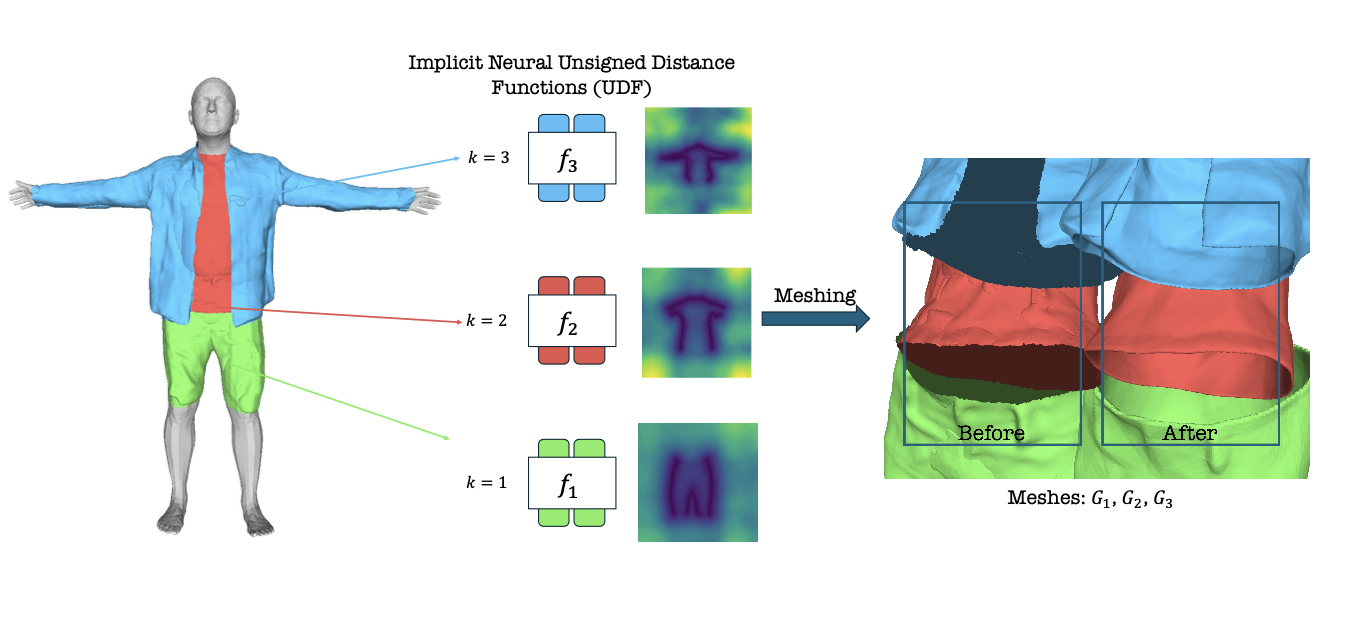}
\caption{\textbf{Visualization of garment refinement}. To mitigate the local geometry errors caused by 3D segmentation (visible on all garment boundaries on the right) and penetration removal (most visible on the red garment on the right) steps, we utilize the power of neural implicit unsigned distance fields. We fit a neural network for each garment layer and construct meshes. }
\label{fig:fit}
\end{figure*}

Previous stages generate penetration-free multi-layered garment areas that can be draped onto a common human body $B_c$. The resulting garments may have two small undesired geometry artifacts; The first type of geometry errors is segmentation errors, which are most evident in the borders. The second type of geometry errors is the potential artifacts that may be generated due to the discontinuous nature of the penetration removal algorithm. To solve both types of errors, we aim to use the fitting power of implicit neural unsigned distance functions. Implicit neural UDFs can approximate the geometry at hand with the desired level of smoothness. This allows the removal of undesired geometry errors. We use \( M_1', \ldots, M_K \) and \( S_1, \ldots, S_K \) to train neural implicit fields \( f_1, f_2, \ldots, f_K \). Later, we convert the UDFs into the garment \( G_1, G_2, \ldots, G_K \).

We train a neural implicit unsigned distance field (UDF) for each layer of the garments. The neural UDFs follow the formulation described in Section~3.1 and are implemented as simple MLPs. Taking inspiration from~\cite{mildenhall2020nerf}, we add \( N \) positional encodings in addition to the 3D point input \( \mathbf{x} \) as described in Eq.~\ref{eq:positional_encoding}. Positional encoding is beneficial because it can provide control over the smoothness and the amount of high-frequency details, as demonstrated by our experiments.

\begin{equation}
PE(\mathbf{x}) =
\begin{cases} 
\sin(2^i\mathbf{x}) & \\
\cos(2^i\mathbf{x}) & 
\end{cases}, \quad i = 0, \ldots, N - 1
\label{eq:positional_encoding}
\end{equation}

We train the neural implicit unsigned distance fields (UDFs) via 3D supervision of points and ground truths. For each \( k = 1, \ldots, K \), we use \( M'_k \) and \( S_k \) to sample points. We divide the sampling process into three categories. The first category is the points that are sampled on the surface of \( M'_k \) and are called on-surface-points. The second category of points is again sampled on the surface of \( M'_k \), but then slightly displaced with an amount sampled from the gaussian \( \mathcal{N}(0, 0.01 \times \mathbb{I}_3) \). The third group of points is sampled on anywhere in a bounding box that contains the mesh \( M'_k \). For all the sampled points, we detect the closest point for them on the surface of \( M'_k \), and the points are filtered out if their closest point does not belong to the set \( S_k \). For the remaining points, we compute their unsigned distance \( d^{gt}_k \) to \( M'_k \) to train the networks described in~\cref{eq:fk}
\begin{equation}
f_k : \mathbb{R}^3 \to \mathbb{R}_{\geq 0}, \quad f_k(x_k) = d_k, \quad k \in \{1, 2, \ldots, K\}
\label{eq:fk}
\end{equation}

To train the UDFs, we use UDF loss from~\cref{eq:loss} with ground truth UDF values \( d^{gt}_k \). We clamp the distance field values with a threshold \(\delta\) to ensure our network focuses near the surface more. After the UDF networks are trained, we apply marching cubes (MC) to convert the garments into triangle meshes.

\begin{equation}
\mathcal{L}_{UDF} = \frac{1}{K} \sum_{k=1}^{K} \mathbb{E}_{x_k \sim S_k} \left[ \| \min(f_k(x_k), \delta) - \min(d_{k}^{gt}, \delta) \|_2^2 \right]
\label{eq:loss}
\end{equation}

\subsection{Double-Sided Mesh with Back Face Normal Orientation}

We use marching cubes with a modest threshold to obtain the double-sided garment with a small thickness. Even though this double-sided mesh is of high quality, to match its look to those of the single-sided surface meshes when rendering, we flip the normals on the back face. We consider the orthogonal camera projection with a viewing direction $d$. For this purpose, we use the joints $J_1, J_2, \ldots, J_{N_j}$ of $B_c$. 

First, we compute the nearest joint $J_{v_i}$ for each vertex $v_i$ that belongs to the double-sided mesh $M$. Then, we look for the projection of $v_i$ on the line $J_{v_i} + td$. If the point is projected sufficiently behind the camera with $t^* > \xi$, we decide that it is a back vertex and its normal should be flipped. After deciding whether normals for each vertex need to be flipped, we flip the faces where all of their vertices require a normal flip. A pseudo-code formulation of this algorithm can be seen in~\cref{alg:face_orientation}.

\begin{algorithm}
\caption{Face Orientation Algorithm}
\label{alg:face_orientation}
\begin{algorithmic}[1]
\Require $M, \{v_1, v_2, \ldots, v_{N_M}\} \in M, B_c, \{J_1, \ldots, J_N\}$
\Ensure $M'$

\Function{OrientFaces}{}
    \State $M' \gets M$
    \State $\text{Flip}[\ ] \gets \text{empty boolean array of size } N_M$
    
    \For{$i \gets 1$ to $N_M$}
        \State $J_{v_i} \gets \text{closest joint of } v_i \text{ from } \{J_1, \ldots, J_N\}$
        \State $l \gets \text{line along viewing direction } J_{v_i} + td$
        \State $t^* \gets \text{projection of } v_i \text{ on } l$
        
        \If{$t^* > \xi$}
            \State $\text{Flip}[i] \gets \text{True}$
        \Else
            \State $\text{Flip}[i] \gets \text{False}$
        \EndIf
    \EndFor
    
    \For{face $f(v_1^f, v_2^f, v_3^f) \in M'$}
        \If{$\text{Flip}[v_1^f]$ and $\text{Flip}[v_2^f]$ and $\text{Flip}[v_3^f]$}
            \State $f(v_1^f, v_2^f, v_3^f) \gets \text{flipped face } f(v_3^f, v_2^f, v_1^f)$
        \EndIf
    \EndFor
    \State \Return $M'$
\EndFunction
\end{algorithmic}
\end{algorithm}

\subsection{Data Capture and Model Design}
For capturing image layers in the wild, we utilize an iPhone 13 Pro Max attached to a tripod. One photo per garment layer needs to be captured with this fixed camera setup. Note that we do not use the camera position or the camera matrix. The images are cropped to a resolution of $1024 \times 1024$ and the backgrounds are removed using Segment Anything~\cite{SAM}.

In the body fitting and garment separation stage, we utilize SMPLer-X~\cite{cai2023smplerx} to obtain human body parameter estimation required by SITH. We use SMPL-X~\cite{SMPL-X} human parametric model to represent the human body. In the garment separation stage, we render multi-view images of the colored watertight meshes of the SITH outputs via rasterization. We utilized 60 circularly sampled views in total. We sampled with $10^{\circ}$ angle separation horizontally and $30^{\circ}$ separation for upper and lower views. Each of the renders is segmented in 2D via GroundedSAM~\cite{ren2024grounded}, and garment masks are obtained. 3D vertices of the SiTH output meshes are projected onto each of these views, and their labels are assigned according to whether their projection falls into the garment mask in the majority of the views.

In the garment penetration removal stage, we utilized a thickness of 2mm for all the garments and utilized NVIDIA kaolin for mesh queries. Garment refinement implementation used N = 4 positional encodings and a 4-layer MLP with hidden dimensions of 128, 256, 256, 128. Rectified linear unit (ReLU) activation was preferred for non-linearity. The number positional encoding is chosen to ensure sufficiently refined garments without damaging the quality of the wrinkles. Finally, our training recipe utilized ADAM  optimizer with a learning rate of $10^{-4}$. To convert our neural implicit UDFs to final garment meshes, we use Marching Cubes with a small threshold of 3mm.

\section{More Experimental Results}

\begin{figure*}[t]
\centering
\includegraphics[width=0.8\linewidth]{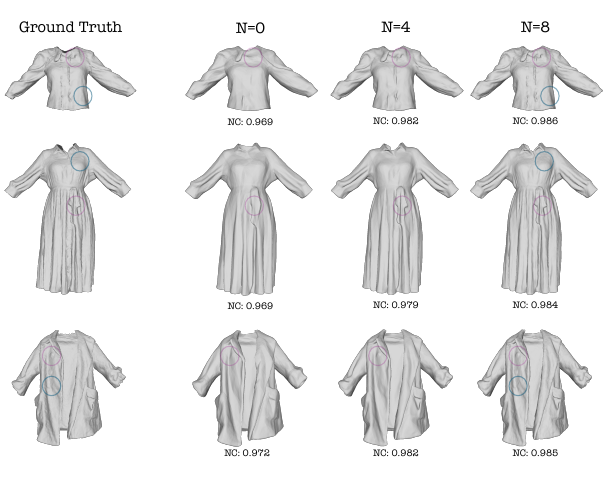}
\caption{\textbf{The number of positional encodings for neural UDFs}. Results for reconstructing the ground truth meshes in the first column after fitting a UDF representation with different numbers of positional encodings and meshing with MC. The last column indicates that our reconstruction can even reconstruct tiny details encircled by blue. Moreover, the first and second columns demonstrate that decreasing the number of positional encodings provides control over the amount of high-frequency details.}
\label{fig:pe}
\end{figure*}

\subsection{Ablation of Positional Encodings for Neural UDF}

In this ablation study, we showcase the results for various complex garments with high-frequency features, where the neural implicit UDFs are trained with different numbers of positional encodings. We train the networks with the same formulation by only changing the number of frequencies in the positional encoding. We utilize Marching Cubes to extract meshes from UDFs with a thickness of 3mm, as we are most interested in modeling the garments with a thickness for our application. We included only normal consistency (NC) but omitted CD, as it was not statistically significant and closely followed the NC results. Here, it is evident that adjusting the frequencies in positional encoding introduces a control mechanism for capturing high-frequency details in the reconstruction. One can see the overly sharp features and missing wrinkles on the reconstruction for N = 0. This means that reconstruction can only be accurate up to an extent when no positional encoding is used. When we increase the number of positional encodings to N = 4, we obtain a much detailed reconstruction with significantly better wrinkle quality and without the unrealistic sharp discontinuities observed for N = 0. We show that with N = 8 frequencies, it is possible to capture all the small details, even the very small blobs that we encircled with blue. However, this means that all the noisy features in the ground truth meshes are also present in the reconstruction. 

\subsection{In-the-wild Reconstruction Results}
We visualize our reconstructed 3D clothing captured with a single iPhone and tripod. In this example, we reconstructed 4 layers of garments on the same human subject.

\begin{figure*}[t]
\centering
\includegraphics[width=\linewidth]{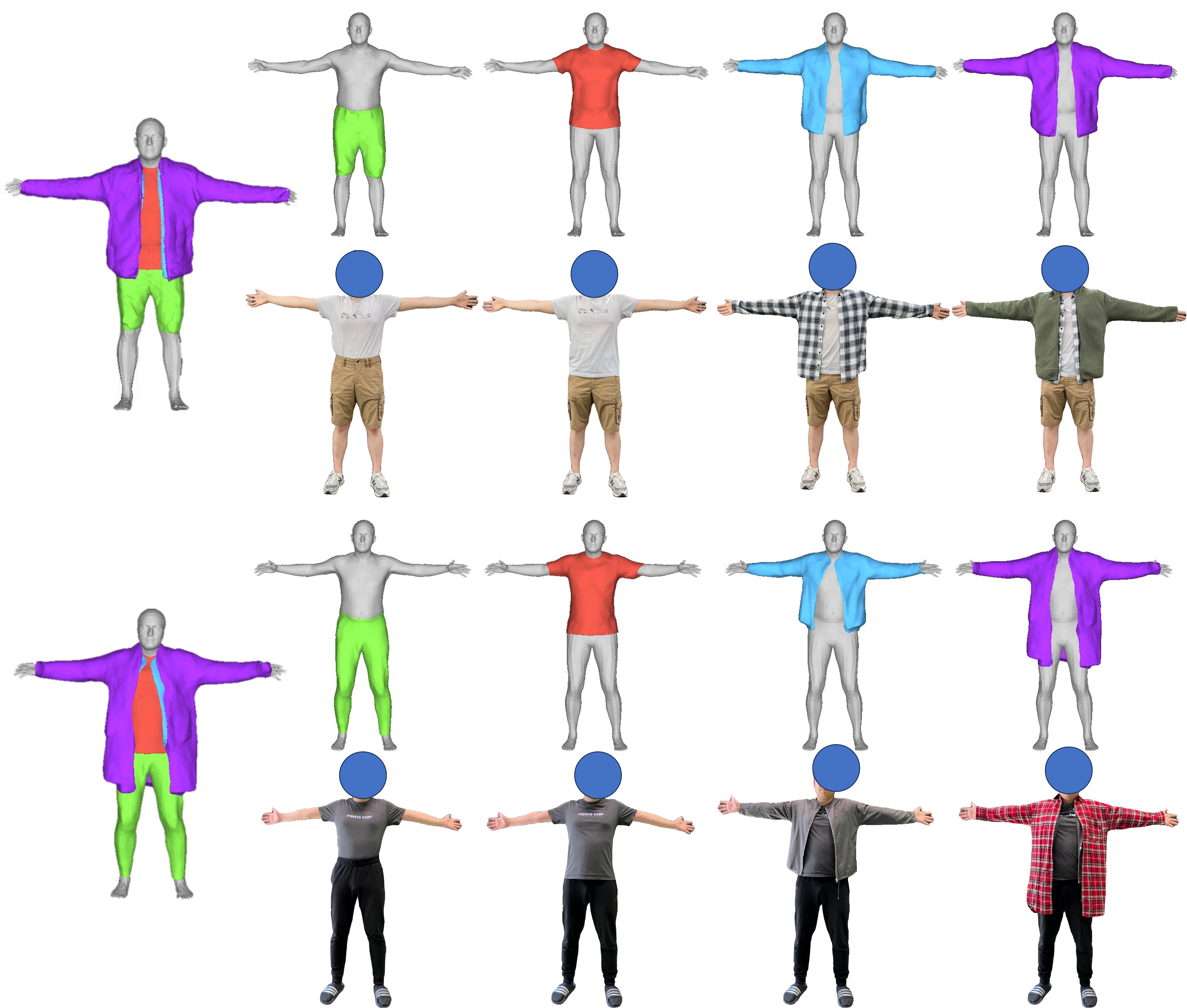}
\caption{\textbf{In-the-wild reconstruction results}. We visualize our reconstructed 3D clothing captured with a single iPhone and tripod. In this example, we reconstructed 4 layers of garments on the same human subject.  }
\label{fig:example}
\end{figure*}

\end{document}